\documentclass{elsart}
\usepackage{epsfig}
\begin{document}
\runauthor{Grzegorz Litak}

\begin{frontmatter}
\title{ Charge and Phase Fluctuations in Attractive Hubbard Model}

\author{Grzegorz Litak\thanksref{E-mail}}
\address{Department of Mechanics, Technical University of Lublin,
Nadbystrzycka 36, PL-20-618 Lublin, Poland}

\thanks[E-mail]{Fax: +48-815250808; E-mail:
litak@archimedes.pol.lublin.pl}

\begin{abstract}
Using the negative $U$ Hubbard model we analyze normal state properties of 
a superconductor. In this model there exists a characteristic pairing 
temperature $T_P$ above a superconducting critical temperature. 
Below $T_P$ electrons start to form incoherent pairs. The fluctuations in 
charge and phase are precursors of charge density wave and 
superconductivity phases depending on band filling. They lead naturally to pseudogap 
opening in the density of states. 
\end{abstract}
\end{frontmatter}

The pseudogap in electronic spectra of HTc superconducting cuprates  has 
attracted  a lot of attention last time \cite{Timusk99}.
The discussion on its origin and nature is not closed \cite{Domanski01} however
one of possible explanations could be that it is a precursor of 
superconducting gap formation \cite{Domanski01,Tobijaszewska00}. In this paper we will follow
that way 
analyzing the phase fluctuations of superconducting order \cite{Gyorffy91} parameter
mediated by fluctuations of charge.  

In this paper we employ the simplest purely electronic model which can lead to 
superconductivity, namely, the negative $U$
Hubbard model \cite{Micnas91}: 
\begin{equation}
  \hat{H} = \sum_{ij\sigma}  
\left(- \mu\delta_{ij}  
 - t_{ij} \right) \hat{c}^+_{i\sigma}\hat{c}_{j\sigma} \nonumber \\
 +\frac{1}{2} U \sum_{i}
 \hat{n}_{i\sigma}\hat{n}_{j-\sigma}, \label{hubbard} \nonumber
\end{equation}
where $i$ and $j$ label the sites
of a  square lattice,
$t_{ij}$
are
electron hopping integrals between nearest neighbour sites,  
$U < 0$ describe attraction
between electrons occupying the same site $i$, $\mu$ is the chemical 
potential.
\begin{figure}[h]
\epsfig{file=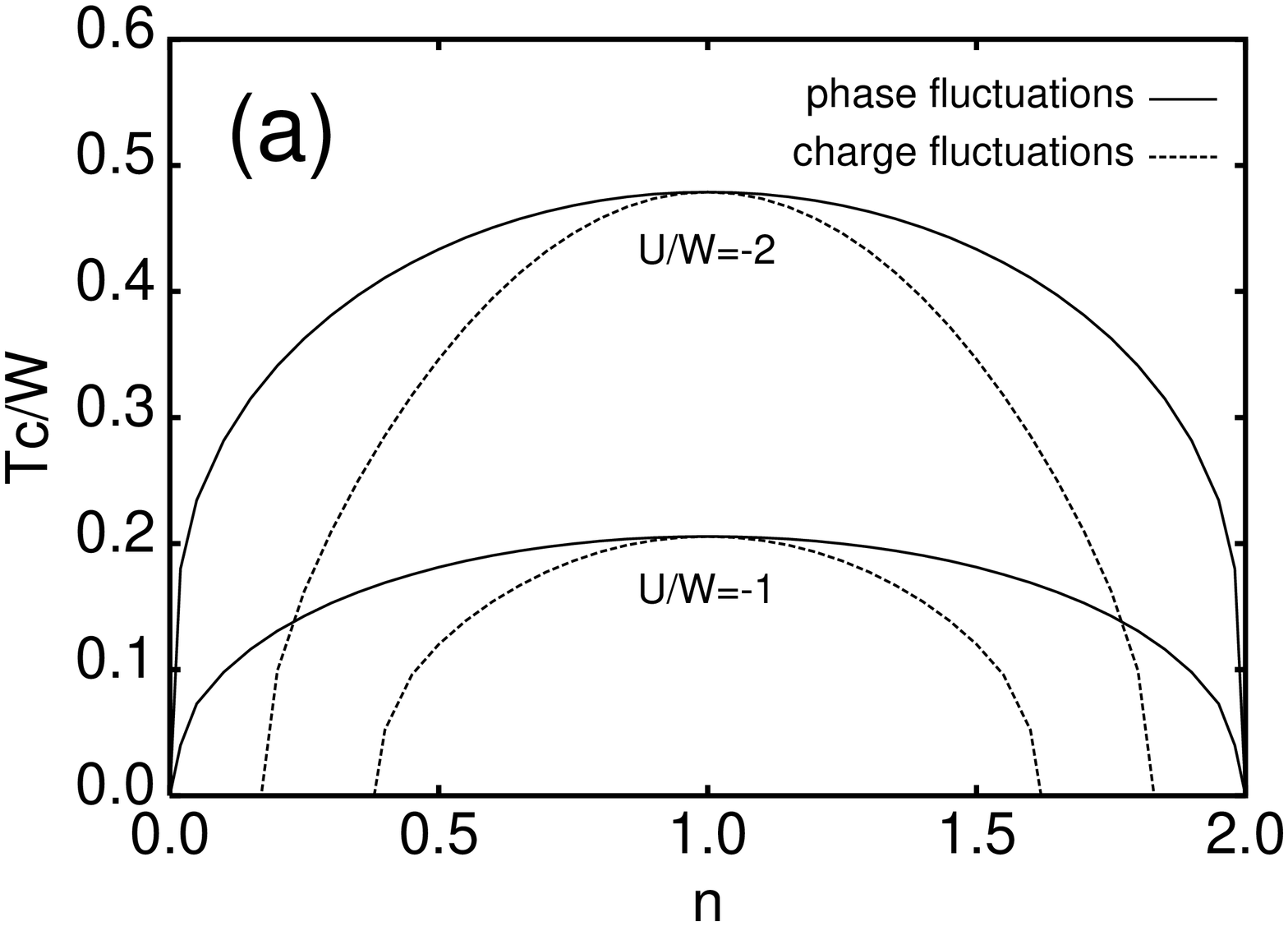,width=5.8cm,angle=-90} \hspace{-1.0cm}
\epsfig{file=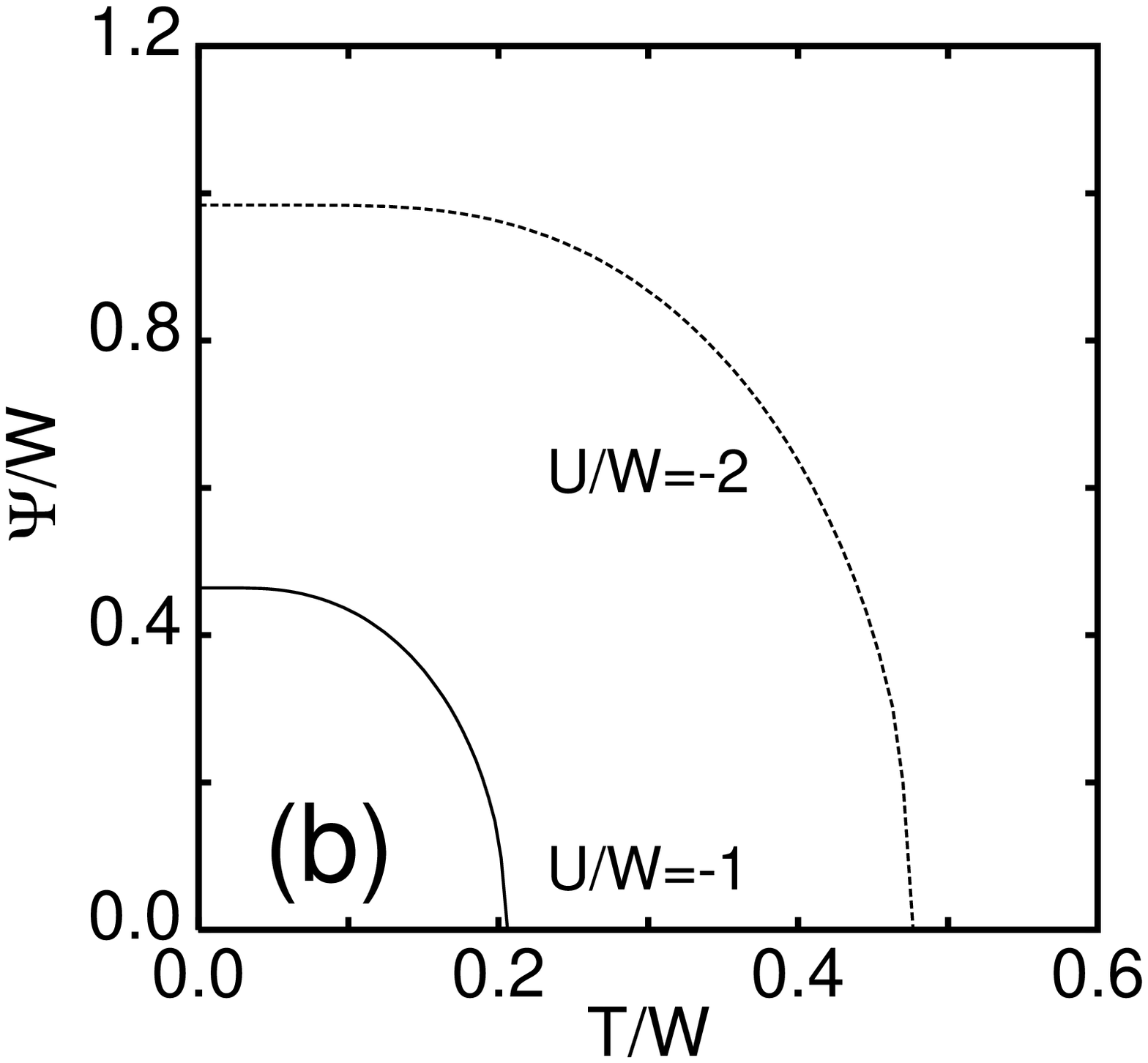,width=5.9cm,angle=-90}
\caption{ (a) Characteristic  temperature $T_C$ for short range
ordering:
local phase and
charge local fluctuations versus band filling $n$,
where
$W=8t$ is a band width. (b) Short range orderer parameter $\Psi$
for half filled band ($n=1$).} 
\label{fig1}
\end{figure}
The self-consistent
Hartree-Fock-Gorkov (HFG) equation:
\begin{equation}
 \sum_{j} \left(\begin{array}{cc}
 (E  +\mu -\frac{U n_i}{2}) \delta_{ij}+ t_{ij} &
 \Delta_{i} \delta_{ij}\\
\Delta_{i}^* \delta_{ij} &  (E  +\mu +\frac{U n_i}{2}) \delta_{ij}- t_{ij}
\end{array}\right)  {\rm \bf G}(j,k;E) = {\bf 1}\delta_{ik}.
\end{equation}
In this case the superconducting order parameter   $\Delta_i$ and the local charge
$n_i$ at a finite temperature $T$ ($\beta=1/kT$) are given by following relations:
\begin{eqnarray}
\Delta_i &=& -\frac{U}{\pi}  \int_{-\infty}^{\infty} {\rm d} E~~ \frac{{\rm
Im}
G^{12}
(i,i; E)}{{\rm e}^{\beta E}+1},~~~
n_i = -\frac{2}{\pi} \int_{-\infty}^{\infty} {\rm d} E~~  \frac{{\rm Im}
G^{11}
(i,i;E)}{{\rm e}^{\beta E}+1}.
\end{eqnarray}
To go beyond the HFG
we apply fluctuations of phase \cite{Gyorffy91} and  charge via
Hubbard III
approximation  \cite{Hubbard64}. Thus random local pairing potential gains 
different phase 
$\Theta_i$, $\Delta_i 
\rightarrow \Delta_i (\Theta_i)=   
|\Delta|
{\rm e }^{i
\Theta_i}$  dependent on lattice site $i$. 
The charge also changes randomly $n_i  \rightarrow  n_i=
n \pm
\delta n_i$ with  lattice site $i$.
\begin{figure}[bh]
\epsfig{file=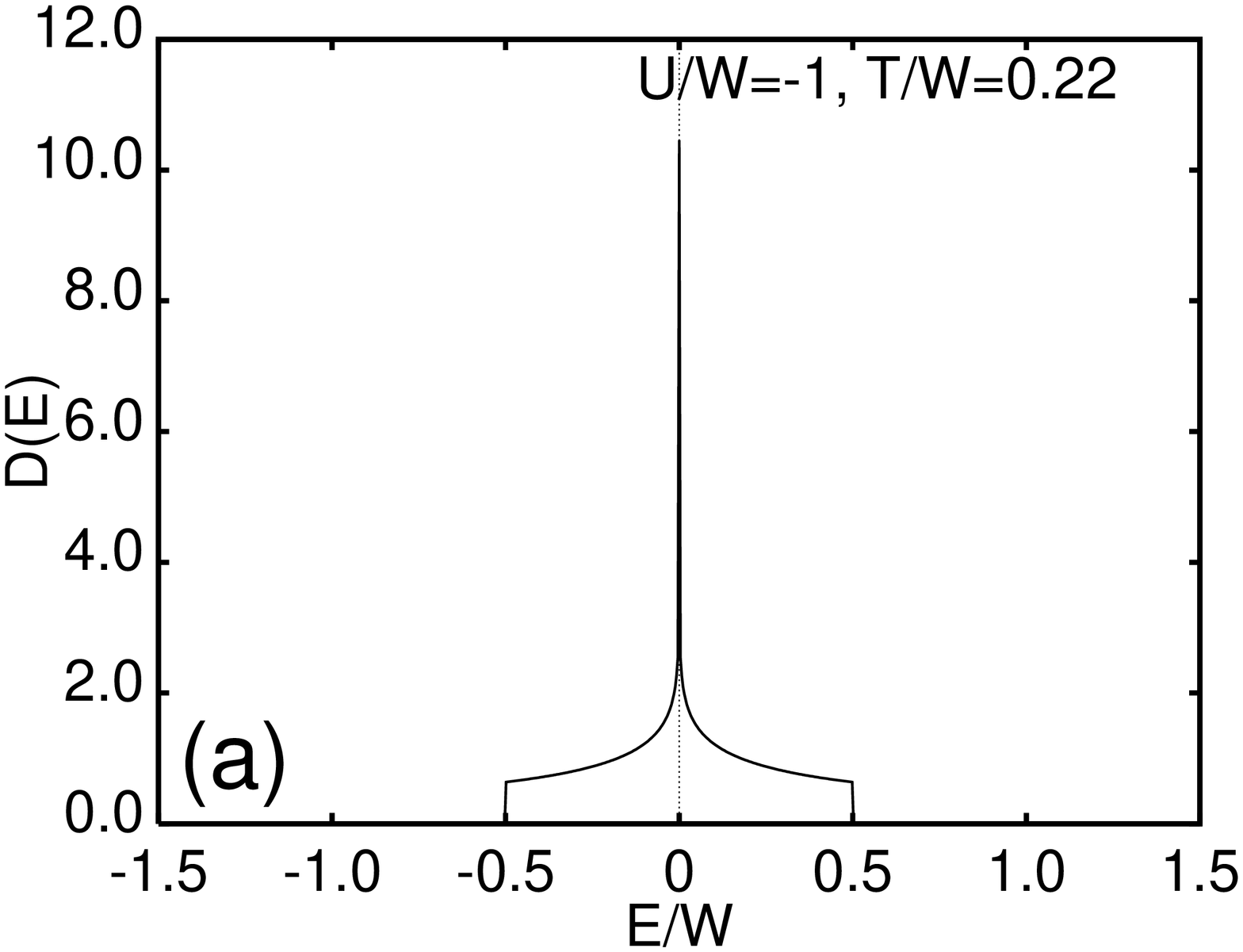,width=5.0cm,angle=-90} \hspace{-0.5cm}
\epsfig{file=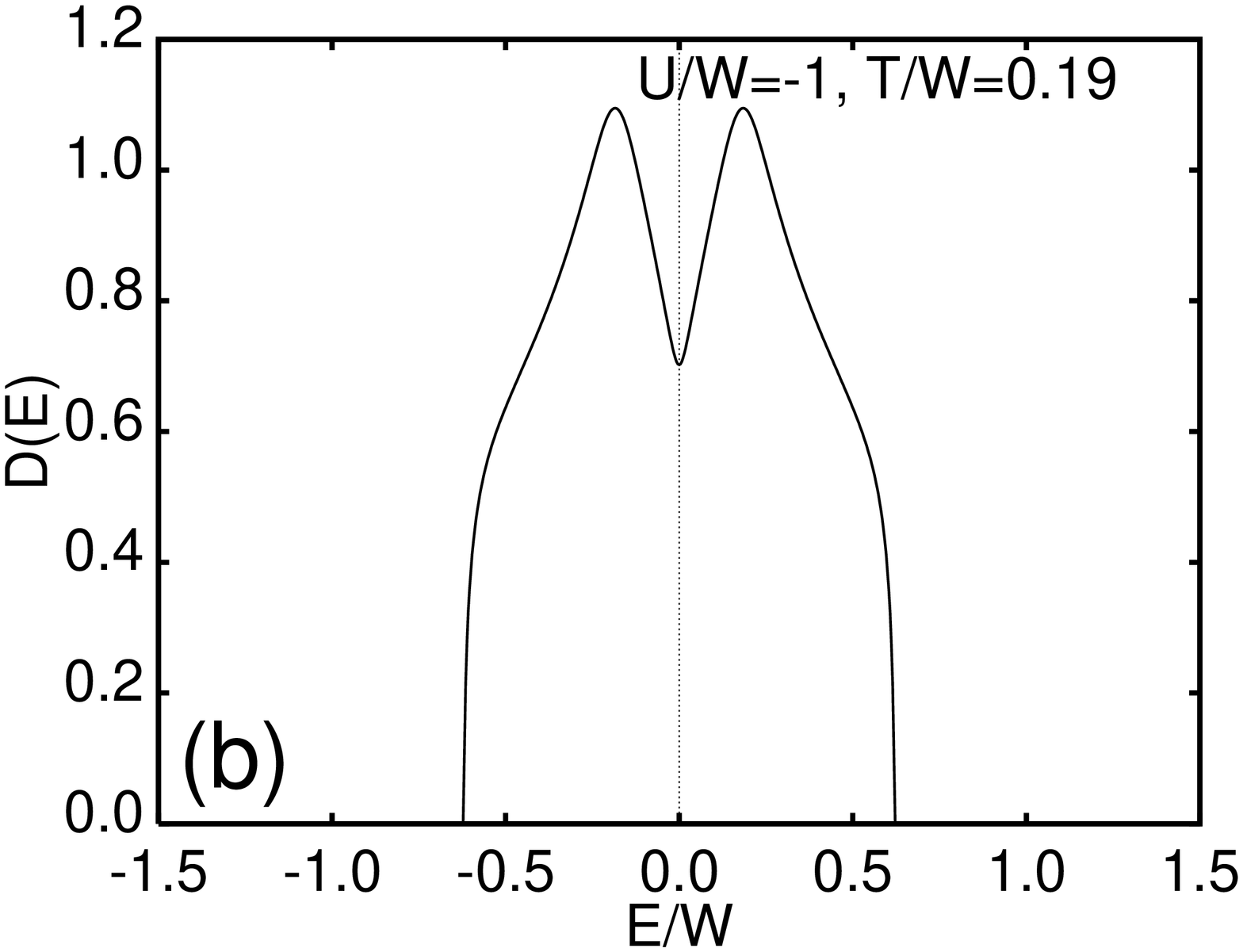,width=5.0cm,angle=-90}

\epsfig{file=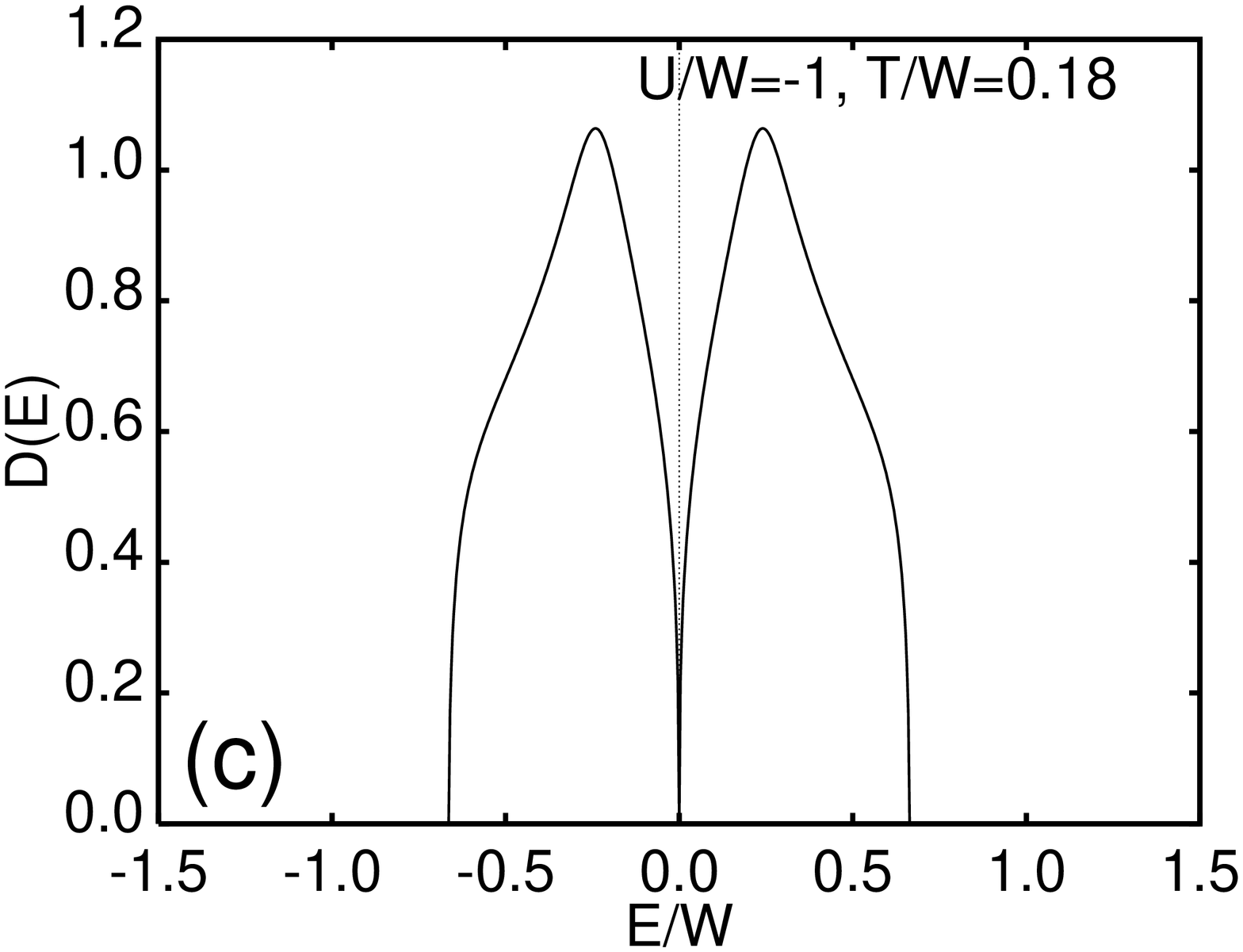,width=5.0cm,angle=-90} \hspace{-0.5cm}
\epsfig{file=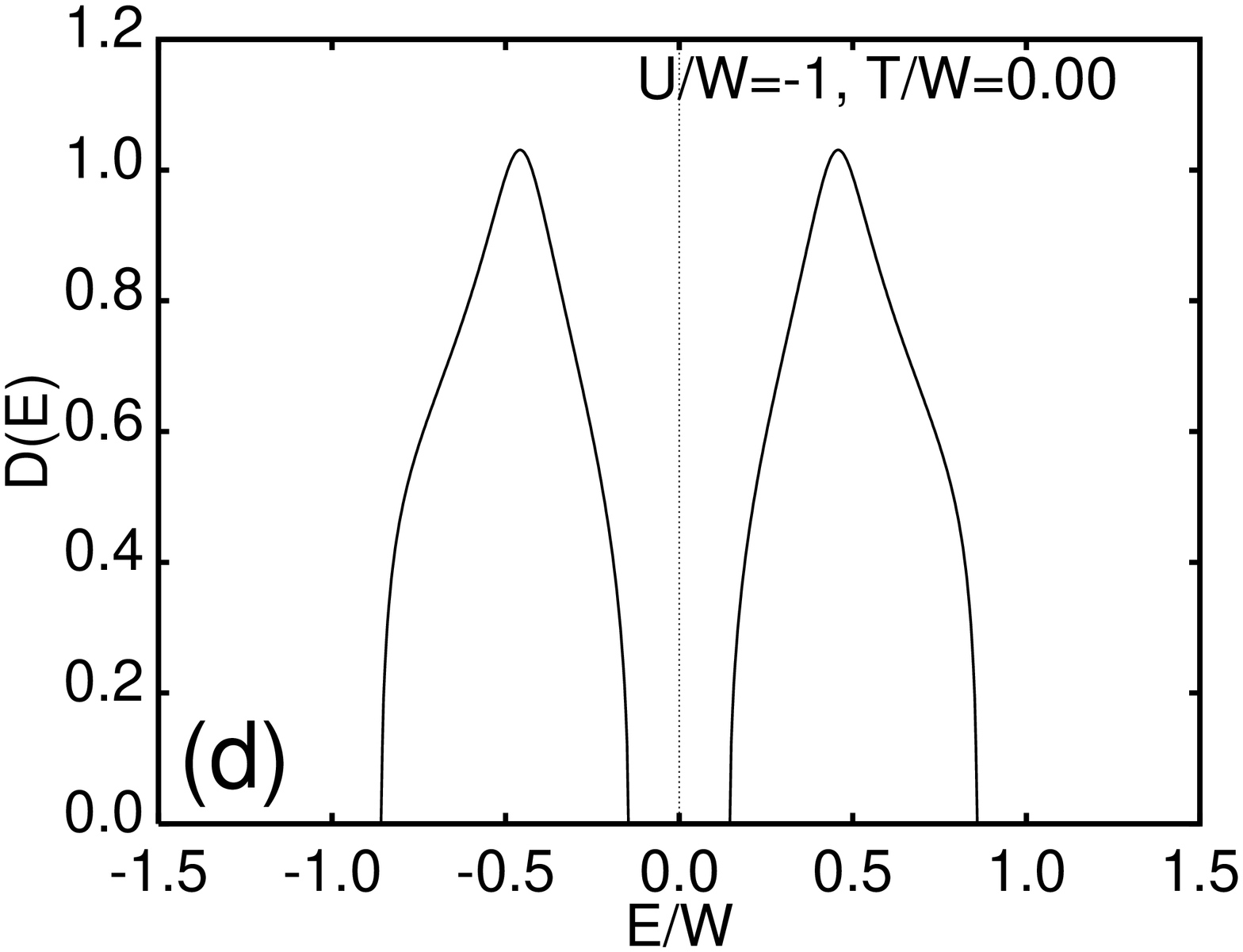,width=5.0cm,angle=-90}
\caption{ Average densities at different temperatures and $n=1$. Note, the
formation
of a pseudogap and a proper  gap is caused by fluctuations in charge and phase
below the critical temperatures of superconductivity and charge density
wave (CDW). } \label{fig2}
\end{figure}
In such approach the equation of motion has the following form:
\begin{equation}
 \sum_{j} \left(\begin{array}{cc}
 (E  +\mu -\frac{U n_i}{2}) \delta_{ij}+ t_{ij} &
 { |\Delta| {\rm e }^{{\rm i} \Theta_i} \delta_{ij}	}\\
 { |\Delta| {\rm e }^{-{\rm i} \Theta_i} \delta_{ij}} &  (E  +\mu 
+\frac{U n_i}{2}) 
\delta_{ij}- t_{ij}
\end{array}\right)  {\rm \bf G}(j,k;E) = \delta_{ik}.
\end{equation}
Here the local pairing parameter   $\Delta_i$ and the local charge
$n_i$ are given by
self--consistent relations:
\begin{eqnarray}
\Delta  ({\Theta_i})  &=& \frac{-U}{\pi}  
\int_{-\infty}^{\infty} {\rm d} 
E~~ 
\frac{ {\rm Im}
G^{12}_{\Theta_i}
(i,i;E)}{{\rm e}^{\beta E}+1},  
\end{eqnarray}
\begin{eqnarray}
 {\delta n} = \frac{n_{ {A} }
-n_{ {B} }}{2}&=& -\frac{1}{\pi} 
\int_{-\infty}^{\infty} {\rm 
d} \omega~~  \frac{{\rm Im}
(G^{11}_{A} (i,i;E))-(G^{11}_{B}
(i,i;E))}{{\rm e}^{\beta E}+1},
\end{eqnarray}
where $\overline \Delta_i =0$ and $\overline n_i= n$ ($ n_A=n+\delta n$ 
and $n_B=n- \delta n$). \\
The set of above equations (Eqs. 4-6) can be solved by means of Coherent 
Potential Approximation
(CPA) \cite{Gyorffy91,Litak92}. Single site condition for the coherent potential 
${\bf 
\Sigma}(i,i, E)={\bf
\Sigma}(E)$
is defined by the zero value of an average  T-matrix:
\begin{equation}
 <{\bf T}_{\alpha,\Theta}(i,i;E)> = <({\bf V}_{\alpha, \Theta} -{\bf 
\Sigma}(E))({\bf 
1}-[{\bf V}_{\alpha, \Theta} 
-{\bf \Sigma}(E)]
\overline{\bf G} (i,i;E))^{-1}>,
\end{equation}
where
\begin{equation}
 V_{\alpha,\Theta} = \left[ \begin{array}{cc} \frac{U n_{\alpha}}{2} & -|\Delta| 
{\rm e }^{{\rm i} \Theta} 
\\ -|\Delta| {\rm e }^{-{\rm i} \Theta}
& -\frac{U n_{\alpha}}{2} 
\end{array} \right]~~~~{\rm and}~~~\alpha=A,B,~~~\Theta \in [0,2\pi]. 
\end{equation}
Equations (5) and (6) can now be expressed through conditionally averaged Green functions:
\begin{equation}
{\bf G}_{\alpha,\Theta} (i,i;E))=\overline{\bf G} (i,i;E)({\bf
1}-[{\bf V}_{\alpha, \Theta}
-{\bf \Sigma}(E)]  
\overline{\bf G} (i,i;E))^{-1}.
\end{equation}

The above  model (Eqs 4-9) has been solved for a square lattice and large 
enough interaction 
$|U|/W=1,2$ ($W=8t$ is a band width).
Fig. 1a shows  characteristic temperatures $T_C$ of phase
(pairing temperature - $T_P$) and charge fluctuations 
appearance. 
Such fluctuations  can
be naturally
related to precursors of long range order phases formation like
superconductivity
and charge density wave (CDW), respectively.
 Note that, fluctuations are characterized    by different
regions 
of  band filling $n$. Charge fluctuations appear around half filled band ($n=1$) while
phase fluctuations are present for any $n$. 
Interestingly, for particle-hole symmetric situation  
 ($n=1$) the  characteristic temperatures $T_C$ are the same for 
both fluctuations and they can coexist (Fig. 1a).
In this case the state can be described by a
generalized local order parameter
$\Psi_i=  \sqrt{ |\Delta|^2+ \left(U \delta n \right)^2}$ (Fig. 1b).
It is similar to low temperature
behaviour
where the coexistence of superconductivity and CDW 
\cite{Miller93}.
However, in that case the CDW state was not stable in presence of diagonal non-magnetic 
disorder \cite{Huscroft97,Litak98,Pawlowski01}.  
The opening of a pseudogap above the superconducting critical temperature
$T_C(sup)$ \cite{Timusk99}
can be easily explained by such fluctuations.
The corresponding densities of states for $U/W=-1$ are 
presented in  Fig. 2. Note that, for smaller temperature $T < 0.18W$  (Fig. 2c) 
the  
pseudogap eluate into a proper gap in electronic spectrum. However to investigate 
further transition into superconducting state one has to investigate Kosterlitz-Thouless  
scenario \cite{Tobijaszewska00}.

{\bf Acknowledgements} \\ 
This work has been partially supported by KBN grant No. 5P03B00221.

\end{document}